\documentclass[11pt,a4]{article}
\usepackage{graphicx}
\usepackage{natbib}
\usepackage{color}
\begin{document}

\title{Anisotropic clustering of inertial particles in
homogeneous shear flow}

\author{P.~Gualtieri, F.~Picano, C.M.~Casciola \\ \\
Dipartimento di Meccanica e Aeronautica,  \\
Universit\`a di Roma {\em{La Sapienza}}  \\
Via Eudossiana 18, 00184 Roma {\it{Italy}} }
\date{}
\maketitle
%___________________________________________________________________________
\begin{abstract}
%_______________________________________________________________________________
Recently, clustering of inertial particles in turbulence has been thoroughly 
analyzed for statistically homogeneous isotropic flows.
Phenomenologically, spatial homogeneity of particles configurations is broken
by the advection of a range of eddies determined by the Stokes relaxation time 
of the particles which results in a multi-scale distribution of local 
concentrations and voids.
Much less is known concerning anisotropic flows.
Here, by addressing direct numerical simulations (DNS) of a statistically 
steady particle-laden homogeneous shear flow, we provide evidence that the
mean shear preferentially orients particle patterns.
By imprinting anisotropy on large scales velocity fluctuations, the shear 
indirectly affects the geometry of the clusters.
Quantitative evaluation is provided by a purposely designed tool, 
the angular distribution function of particle pairs (ADF), which allows
to address the anisotropy content of particles aggregates on a scale by scale 
basis.
The data provide evidence that, depending on the Stokes relaxation time of the 
particles, anisotropic clustering may occur even in the range of scales
where the carrier phase velocity field is already recovering isotropy.
The strength of the singularity in the anisotropic component of the ADF
quantifies the level of fine scale anisotropy, which may even reach values 
of more than $30 \%$ direction-dependent variation in the probability to 
find two close-by particles at viscous scale separation.
%_______________________________________________________________________________
\end{abstract}
%___________________________________________________________________________
\section{Introduction}
%_______________________________________________________________________________
Transport of inertial particles is involved in several fields of science,
e.g. droplets growth in clouds, \cite{falfouste,shaw}, planetary 
formations, \cite{braccoetal}, or plankton accumulation in the 
ocean, \cite{karetal}.
As far as technological applications are concerned, inertial particle dynamics 
is crucial for solid or liquid fuelled rockets, injection systems
of internal combustion engines or for sediments accumulation in pipelines, 
e.g.~\cite{roueat,marsol}.
Inertial particles differ from perfectly Lagrangian tracers due to inertia
which prevents them from following the flow trajectories. 
The main effect consists of ``preferential accumulation'', see for 
instance~\cite{squeat,roueat}. In homogeneous isotropic conditions it amounts 
to the small-scale clustering discussed in a number of recent papers, 
\cite{becetal,reacol, balfalfou}.

In presence of inhomogeneity new features emerge leading to the so-called 
turbophoresis as a preferential accumulation near the boundary in  wall 
turbulence \cite{reeks}.
Under appropriate conditions, particles may achieve extremely large
concentrations at the wall with a substantial reduction of mobility.
This turbulence-induced transport and the issuing preferential accumulation as
been addressed in a number of papers dealing with a variety of configurations,
from boundary layers to planar channels and pipes, attacked both from the
experimental \cite{rigrom,kafetal_1,kafetal_2} and the numerical
\cite{roueat,marsol,olipor} side. 
Though a complete understanding of the phenomenon is still lacking, the 
advection of the particles by the coherent motions in the wall-layer is 
certainly essential, as discussed by \cite{hanratty}. In other words the
structures responsible for particle accumulation at the wall are the same
which sustain turbulence kinetic energy production in the buffer layer,
see also \cite{roueat,marsol}.
Turbophoresis and small scale clustering are different aspects of the same 
inertial particle dynamics. 
Both phenomena are induced by non trivial phase relationships due to 
quasi-coherent vortical structures.
The main difference is provided by the characteristic scales, associated with 
the Kolmogorov time unit $\tau_\eta = \eta^2/\nu$ in one case--$\eta$ being 
Kolmogorov length and $\nu$ the kinematic viscosity--and with the larger 
energy producing time scale in the other.
Inhomogeneity is essential to have spatial segregation. For instance, in the
kinetic model presented in \cite{reeks}, the spatial transport of particle 
concentration presents, beside a Fick-like gradient type diffusion component, 
a contribution associated with the spatial variations of turbulence intensity.
However anisotropy is probably a key ingredient of the process, see e.g.
the preferential direction of the trajectories of particles approaching the 
wall \cite{hanratty}.
The two features are strongly entangled in wall bounded flows. 
A special flow exists however--the homogeneous shear flow 
in a confined box--which retains most 
of the anisotropic dynamics of wall bounded flows still preserving, on 
average, spatial homogeneity.

The flow is bound by a computational box of fixed extension
and its integral scale grows initially to eventually saturate due to 
confinement. Target of the analysis is the statically steady state
with time independent ensemble averages. 
Similar features are found in the the experimental data by \cite{shen}.
By using an active grid to generate a flow with integral scale close to the
transversal dimension of the apparatus, the authors were able to achieve
confinement from the outstart. As a consequence, 
the integral scale did not increase downstream, see also the discussion in 
\cite{les_jfm}.

Our flow shares with the wall-layer streamwise vortices and turbulent kinetic 
energy production mechanisms.  In the numerical experiment, this corresponds
to pseudo-cyclic fluctuations associated with the regeneration of streamwise 
vortical structures.

Velocity fluctuations are strongly anisotropic 
at the large scales driven by production while, for smaller separations, the 
classical energy transfer mechanisms become effective in inducing 
re-isotropization.
This classical issue, see e.g.~\cite{Corrsin}, has been recently revisited by 
more complete diagnostic tools (e.g. SO(3) decomposition of turbulent 
fluctuations) able to quantify on a scale-by-scale basis the amount of 
anisotropy in the carrier fluid, as discussed both experimentally 
\cite{warshe,jaccastalalf} and numerically \cite{casguajacpiv} (see 
\cite{bifpro} for a review). 

Despite anisotropy of the velocity field is now well understood and the 
carrier fluid shows tendency towards isotropy recovery below the shear scale, 
$L_{\rm s} =\sqrt{\epsilon/S^3}$ with $\epsilon$ the average turbulent kinetic
energy dissipation rate per unit mass and $S$ the average shear rate, the 
behavior of particle distributions is still not fully explored.

Anisotropic transport of inertial particles has been recently addressed 
by  \cite{shobal}, who analyzed numerically the initial transient of the 
homogeneous shear flow, i.e.  before saturation of the integral scale occurs, 
with the purpose of modeling unconfined conditions.
The focus was mainly on the comparison of different particle dynamics models.
However, by considering particle configurations in orthogonal planes, the 
authors also discussed the anisotropy of particle clusters concluding that 
particles are most concentrated in the streamwise and least concentrated in 
the cross-stream direction.
The same flow was dealt with by \cite{ahmelg_1,ahmelg_2,shomaspan}
to investigate issues such as turbulence modulation in the two way coupling 
regime or heat transfer induced by the disperse phase.

Purpose of the present paper is the quantitative evaluation of the shear
induced anisotropy in particles clustering. It is now well known that particles
respond to the fluid velocity fluctuations in a certain range of scales
which is determined by their Stokes time. The relevant parameter 
is the Stokes number, ratio of particle Stokes time and flow time scale.
In order to work with well defined conditions, one needs a shear flow 
whose characteristic time and length scales are constant in time. 
The best candidate, is the statistically steady homogeneous shear flow in a 
confined box we have described above. 
In this flow, below the shear scale, the velocity fluctuations 
tend to recover isotropy. The question is then what happens to particle 
clusters.  Do they become isotropic in the smallest scales? For 
given velocity field, how is the geometry of the clusters affected by the 
relaxation time of the particles?

In fact, the main contribution of the present study is the quantitative 
evidence that particle distributions do not necessarily reduce their 
anisotropy at small scales, despite the isotropy recovery occurring in the 
velocity field.
Rather clusters anisotropy  may even grow below the Kolmogorov length where 
the velocity field is smooth and almost isotropic.  
As a matter of fact, inertia manifest itself in a rather peculiar and 
unexpected way, and leads, under certain coupling conditions, 
to singular 
particle distributions which viscosity cannot regularize~\cite{becetal}.
After introducing a suitable observable--the angular distribution function-- 
its spherical decomposition is used to evaluate the relative importance of its 
different components. The scaling exponents of the 
respective singularities show that, under appropriate conditions, 
anisotropy is a leading order effect which may easily persist down to 
vanishing separations.  The data offer preliminary evidence of the 
anisotropic geometry of the fractal support of inertial particle distributions 
under shear flows, thus non-trivially extending results recently achieved for
isotropic transport.
%-----------------------------------------------------------------
\section{Methodology}
\begin{figure}
\begin{center}
\includegraphics[width=.33\textwidth]{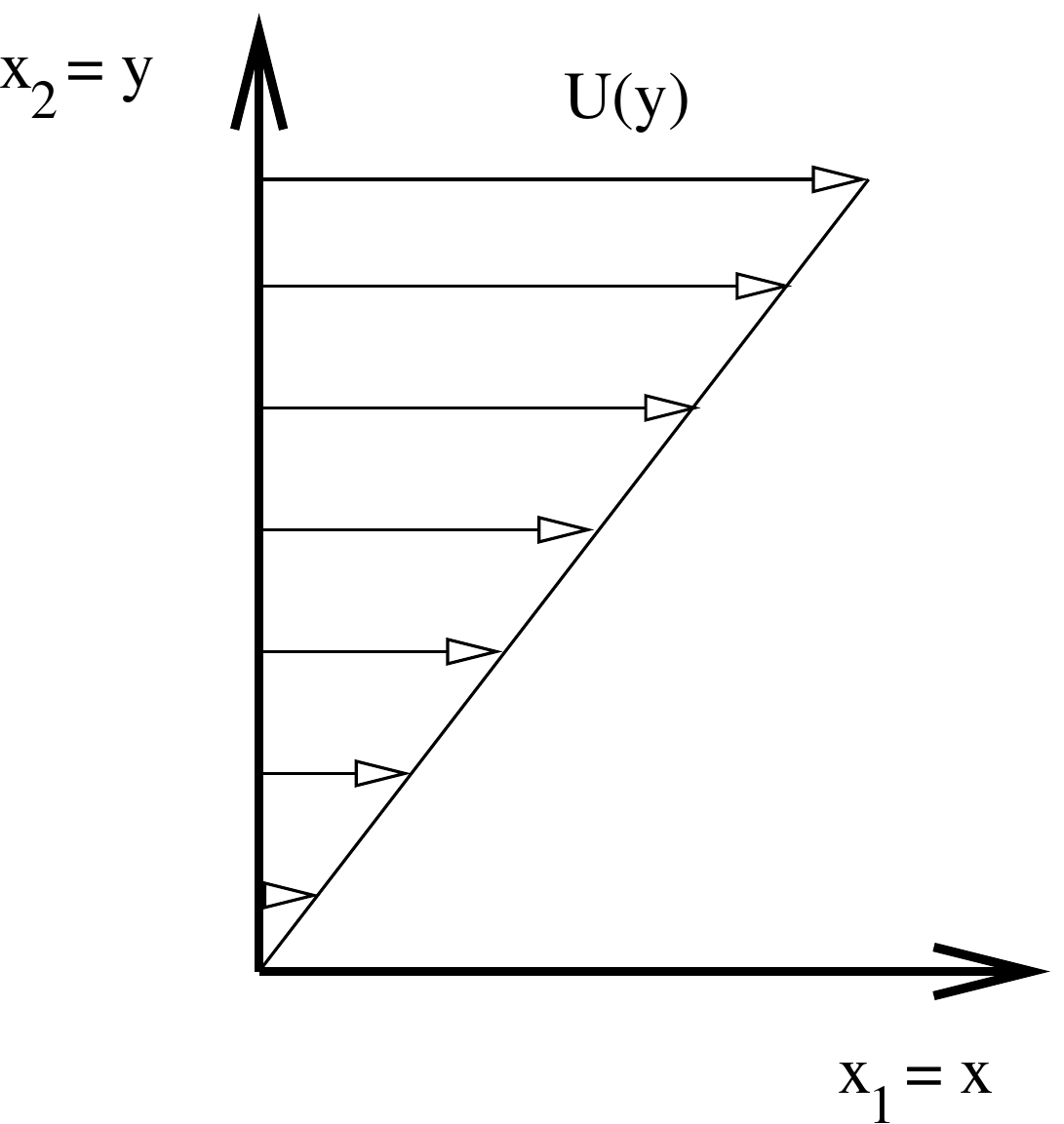} \hfill
\includegraphics[width=.4\textwidth]{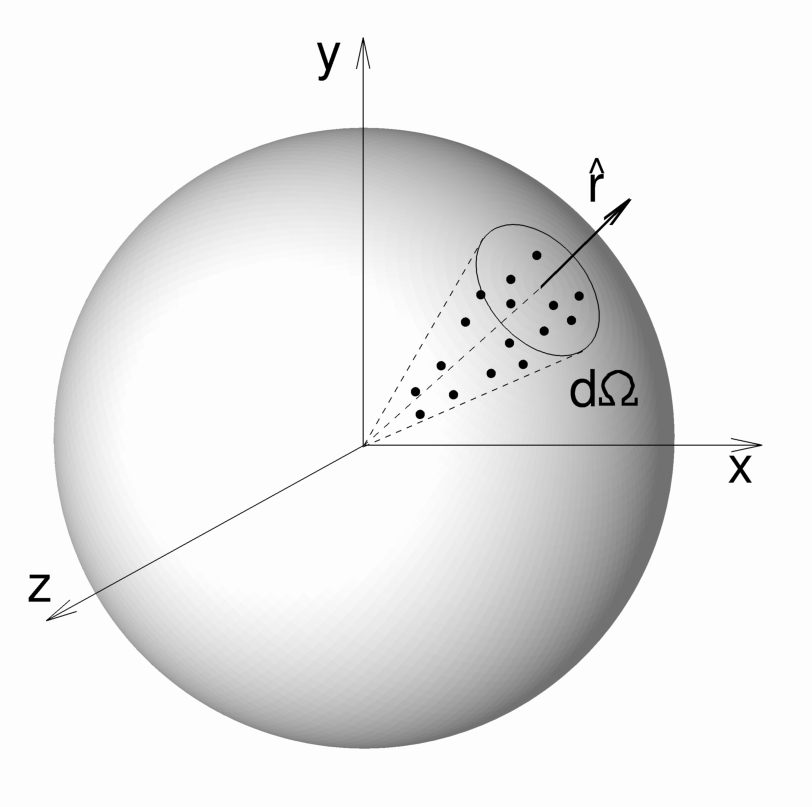}\\
\end{center}
\caption{
Left: sketch of the shear flow and nomenclature: the mean flow $U(y)$,
      in the $x \equiv x_1$ direction, is a function of $y \equiv x_2$,
      with $z \equiv x_3$. For a linear mean profile, the shear rate 
      $S = dU/dy$ is constant. 
Right: sketch of the spherical cone of amplitude $d \Omega$ in direction 
       ${\bf \hat{r}}$.
         \label{fig:sketch}}
\end{figure}

%_______________________________________________________________________________
Concerning the carrier fluid, the velocity field ${\bf v}$  is decomposed into 
a mean flow ${\bf  U}=S x_2 \, {\bf e}_1$ and a fluctuation ${\bf u}$, see 
figure \ref{fig:sketch} for notations. 
Rogallo's technique is employed to rewrite the Navier-stokes equations for 
velocity fluctuations in a deforming coordinate system convected by 
the mean flow according to the transformation of variables 
$\xi_1=x_1 - S t x_2; \quad \xi_2=x_2; \quad \xi_3=x_3; \quad \tau=t$, 
\cite{rogallo}. 
The resulting system is numerically integrated by a 
pseudo-spectral method combined with a fourth order Runge-Kutta scheme for 
temporal evolution, see \cite{guacasbenamapiv}.

The two parameters controlling the homogeneous shear flow are the 
Taylor-Reynolds number
${\rm Re}_\lambda = \sqrt{5/(\nu \epsilon)} \langle u_\alpha u_\alpha\rangle$
and the shear strength 
$S^* =  S \langle u_\alpha u_\alpha \rangle/\epsilon$.
For the simulations discussed below they are 
${\rm Re}_\lambda \simeq 100$ and $S^* \simeq 7$, corresponding to a
ratio of shear to Kolmogorov scale $L_s/\eta \simeq 35$. 
Navier-Stokes equations are integrated in a 
$4\pi\times 2 \pi \times 2 \pi$ periodic box
with a resolution 
of $256 \times 256 \times 128$ Fourier modes corresponding to 
$384 \times 384 \times 192$ collocation points in physical space due
to the $3/2$ dealiasing rule. The Kolmogorov scale is 
$\eta=0.02$ which correspond to $K_{max} \eta=3.1$ ensuring sufficient 
resoltion at small scales. 
Actually a well resolved velocity field is crucial to minimize numerical 
errors associated with the interpolation of the fluid velocity at particle 
positions which is necessary to advect the particles in the present mixed 
Eulerian-Lagrangian formulation, see e.g. \cite{yeung,balachandar}.

\begin{figure}
\begin{center}
\includegraphics[width=.49\textwidth]{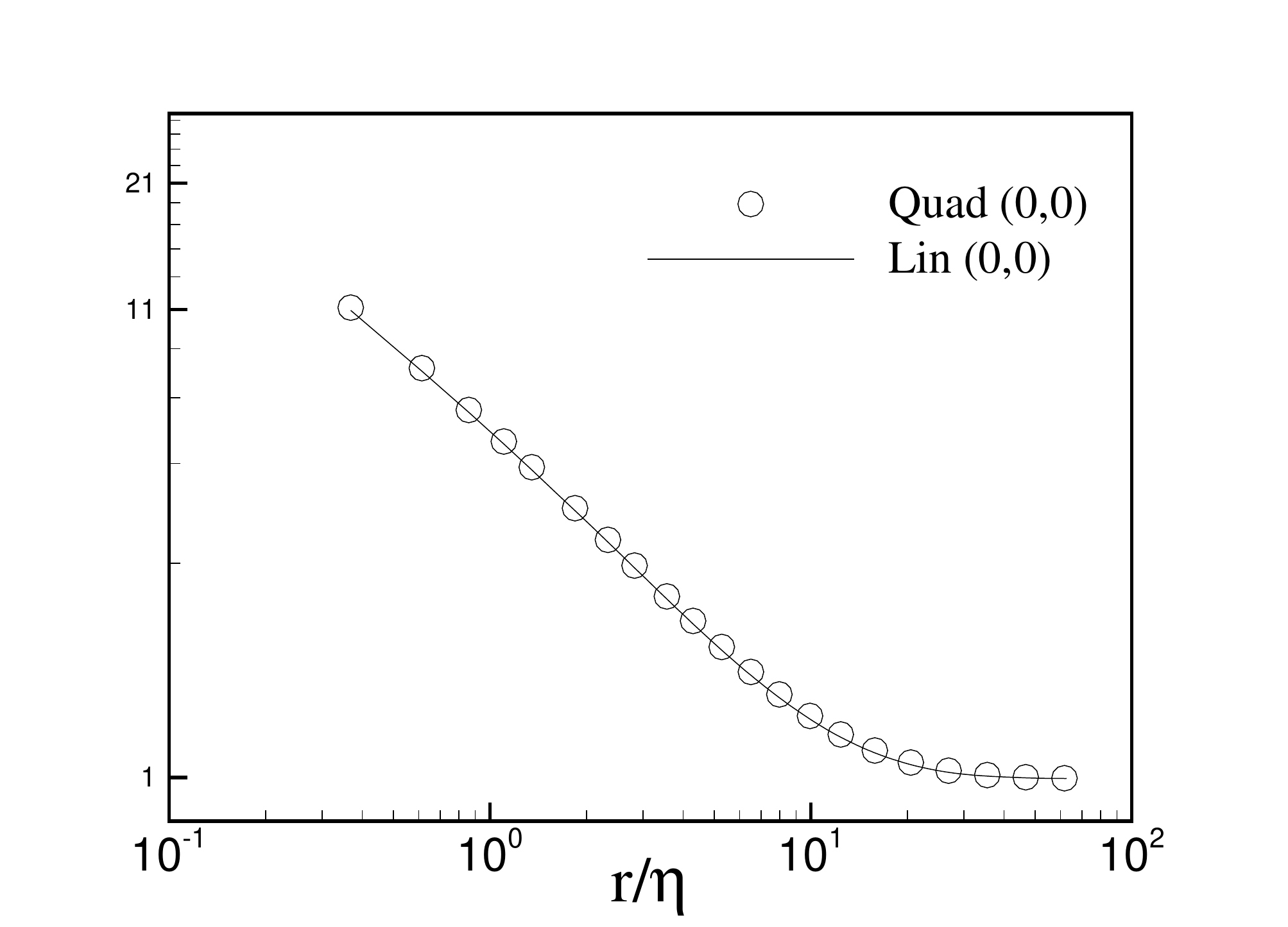}
\includegraphics[width=.49\textwidth]{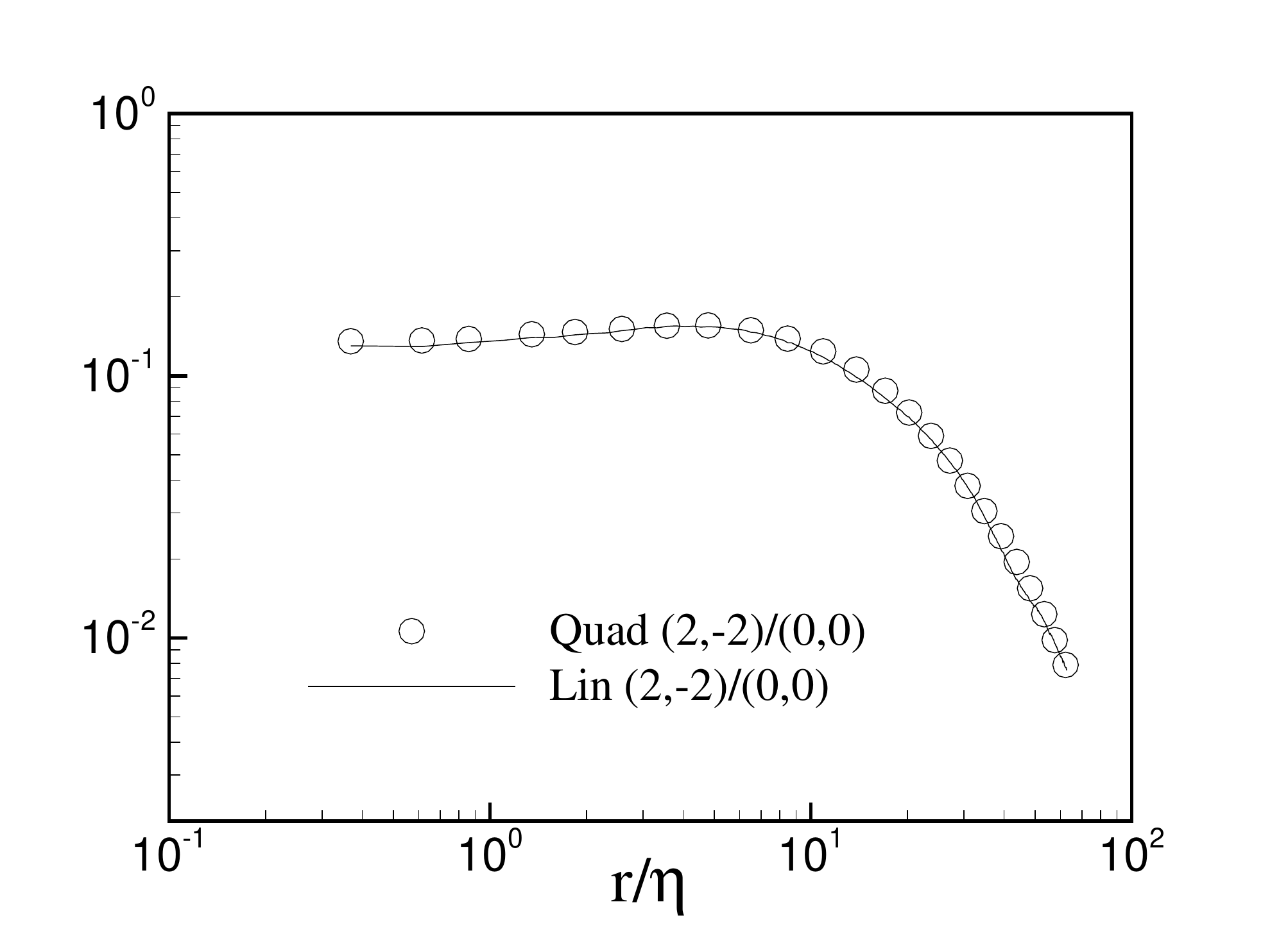}
\end{center}
\caption{DNS of particle laden homogeneous shear flow  
($Re_\lambda=60$, $S^*=7$, $St_\eta=1$), in a 
$4 \pi \times 2 \pi \times 2 \pi$ computational box with
$192 \times 192 \times 96$ collocations points corresponding to
$K_{max} \eta=3$. 
The disperse
phase is computed by using two different schemes to interpolate
the fluid velocity at particle positions, namely linear interpolation 
(solid line) and quadratic Lagrange polynomials (symbols). 
The statistical observables shown in the plots are defined in \S~\ref{results}.
Left panel: $g_{00}(r)$, projection of the ADF in the isotropic sector. 
Right panel: $|g_{2-2}|/g_{00}$, normalized most energetic anisotropic 
component of the ADF.
\label{interp_comp}}
\end{figure}

The disperse phase consists of diluted particles with mass density 
$\rho_p$ much larger than the carrier fluid $\rho_f$, assumed small enough
to be modeled as material points.
At this dilution self-interactions and the back reaction on the
fluid are negligible, leaving the Stokes drag in the relative motion 
with the fluid as the only relevant force on each particle, \cite{maxril}.
Accordingly, the equations for particles position $x^p_i(t)$ and velocity 
$v^p_i(t)$ read
%_______________________________________________________________________________
\begin{equation}
\label{eq_part_phys}
\begin{array}{l}
\displaystyle  \frac{d x^p_i}{dt} = v^p_i \\ \\
\displaystyle  \frac{d v^p_i}{dt} = \frac{1}{\tau_p} 
                \left[ v_i(x^p,t) - v^p_i(t) \right] 
\end{array}
\end{equation}
%_______________________________________________________________________________
where $v_i(x^p,t)$ is the instantaneous fluid velocity evaluated at 
$x^p_i(t)$ and $\tau_p=\rho_p d_p^2/(18 \nu \rho_f)$ is the 
Stokes relaxation time ($d_p$ denotes the particle diameter).
The particle dynamics is controlled by the ratio of $\tau_p$ to a 
characteristic flow time scale, typically the Kolmogorov time scale 
$\tau_\eta=\eta^2/\nu$, i.e. the relevant control parameter is the Stokes 
number $St_\eta=\tau_p/\tau_\eta$.
Particle velocities are decomposed as
$v_i^p=U_i[x_k^p(t)] + u_i^p$ where $u_i^p$ denotes the particle 
velocity deviation with respect to the local mean flow of the carrier fluid.
Eqs. (\ref{eq_part_phys}) can be written as
%_______________________________________________________________________________
\begin{equation}
\label{eq_part_phys_fl}
\begin{array}{l}
\displaystyle  \frac{d x^p_i}{dt} = u^p_i + U_i(x^p)\\ \\
\displaystyle  \frac{d u^p_i}{dt} = \frac{1}{\tau_p} 
                \left[ u_i(x^p,t) - u^p_i(t) \right] - \frac{d U_i}{d t} \, ,
\end{array}
\end{equation}
%_______________________________________________________________________________
to be finally rearranged in Rogallo's computational space as (see also 
\cite{shobal})
%_______________________________________________________________________________
\begin{equation}
\label{eq_part_comp}
\begin{array}{l}
\displaystyle  \frac{d \xi^p_i}{d \tau} = u^p_i - S \tau u^p_2 \delta_{i1} \\ \\
\displaystyle  \frac{d u^p_i}{d\tau} = \frac{1}{\tau_p} 
                \left[ u_i(\xi^p,\tau) - u^p_i(\tau) \right] - S u^p_2 \delta_{i1} \ .
\end{array}
\end{equation}
%_______________________________________________________________________________
The particle equations are integrated by the same fourth order Runge-Kutta 
scheme used for the Navier-Stokes equations, with fluid velocity at particle 
positions evaluated by tri-linear interpolation.
The accuracy of the interpolation scheme  may be an issue.
To assess its effect on the numerical results, we have preliminarily run two 
different simulations at half the resolution and smaller Reynolds number of 
the cases to be discussed in the main body of the paper.
The two simulations employ two different interpolation schemes, namely a 
linear and a quadratic one. 
As always in turbulence, comparisons need to be made in terms of the relevant 
statistical observables.
We anticipate that here we deal with the different projections of the 
Angular Distribution Function, to be introduced in \S~\ref{results}.
In figure~\ref{interp_comp} the solid lines denote results obtained with the 
linear interpolation scheme while symbols correspond to the quadratic Lagrange 
polynomials. 
The difference cannot be  appreciated on the scale of the diagram, and is 
always below the statistical accuracy of the data.

Starting from an already fully developed fluid velocity field in statistically 
steady conditions, five different populations of $N_p=300000$ particles each, 
with Stokes numbers $St_\eta=0.1,\, 0.5,\, 1.0, \, 5.0, \, 10.0$, are 
initialized with random and homogeneous positions and velocities 
matching that of the local fluid. 
Samples for particle statistics corresponding to $120$ independent snapshots
separated in time by $2\,S^{-1}$ are collected after an initial transient 
of $50 \, S^{-1}$. 
Discarding the initial transient is crucial to have results independent from 
the rather arbitrary initial state used to initialize the particles. 
We observe that, also under this respect, a statisticaly steady flow is 
mandatory to have well definite experimental conditions, especially in cases
where the response of particle populations with different relaxation times are 
compared.
%_______________________________________________________________________________
\section{Results \& discussion} \label{results}
\begin{figure}
\begin{center}
\includegraphics[width=.9\textwidth]{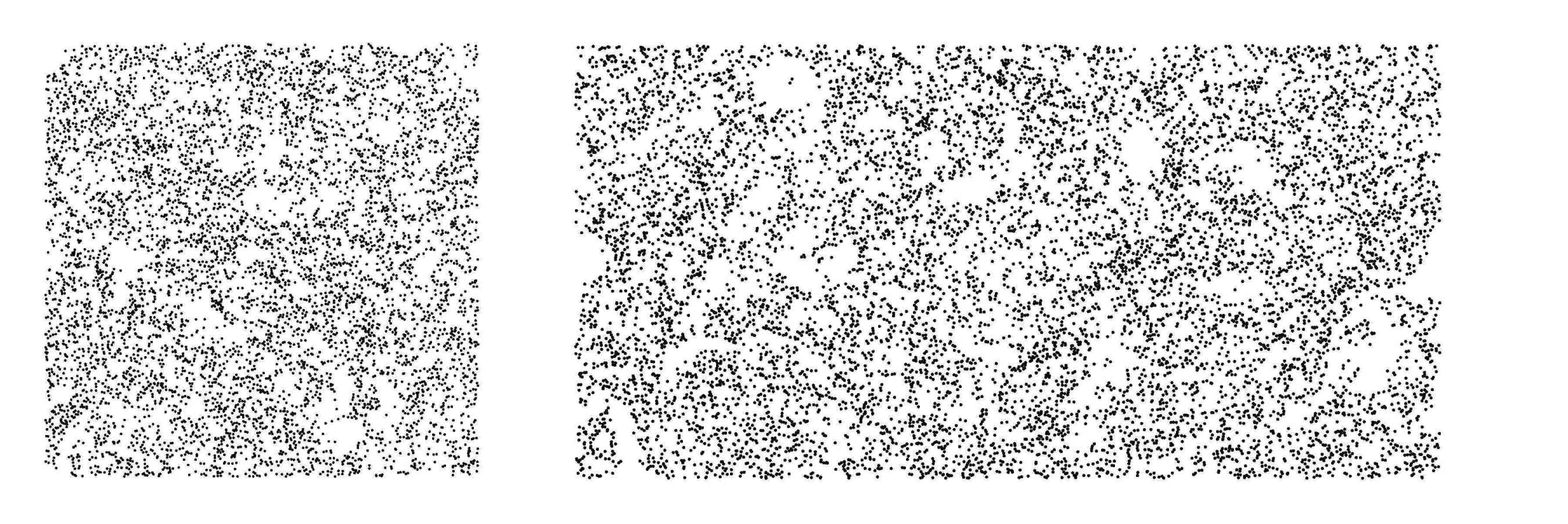}\\
\includegraphics[width=.9\textwidth]{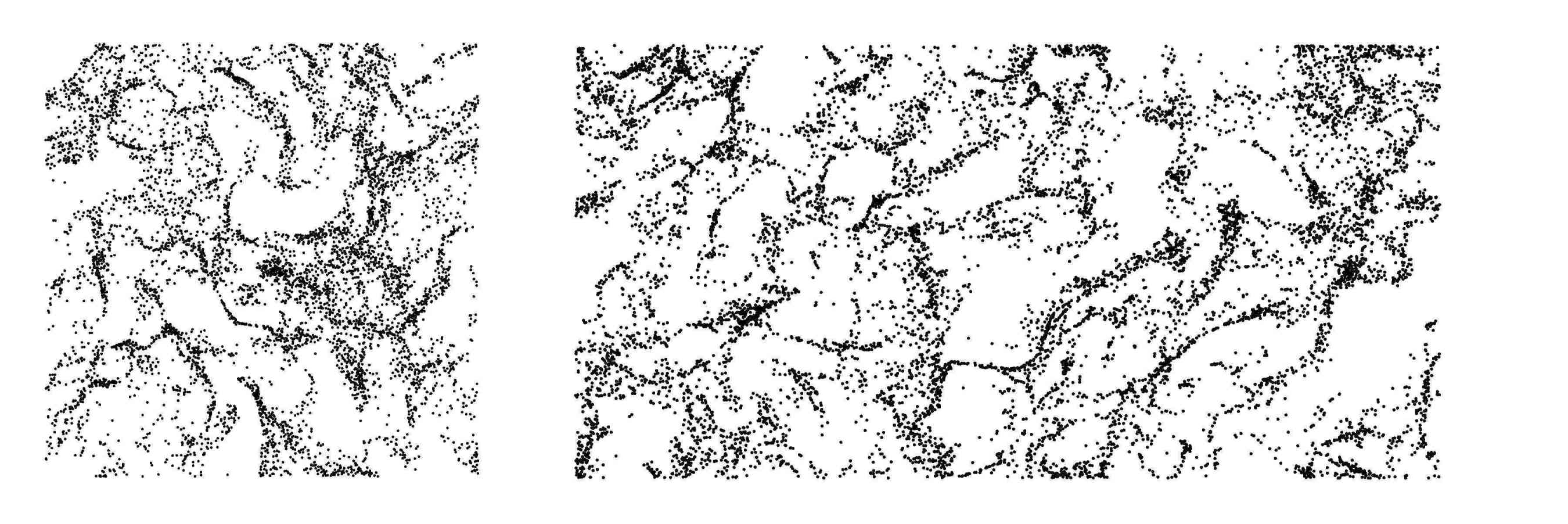}\\
\includegraphics[width=.9\textwidth]{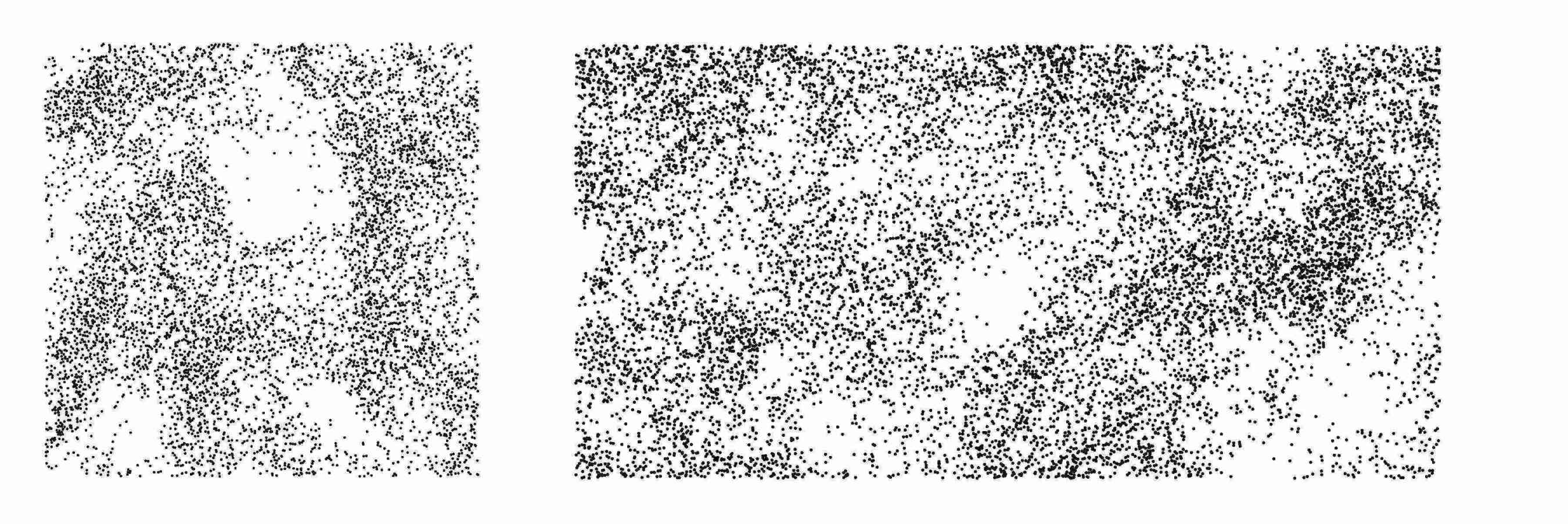}
\end{center}
\caption{Snapshots of particle positions for increasing Stokes number,
from top to bottom: $St_\eta=0.1,\,1,\,10$ respectively. Left column thin slice
in the $y-z$ plane; right column slice in the $x-y$ plane. The
slice thickness is of the order of a few Kolmogorov scales.
         \label{fig:part_pos} }
\end{figure}

%_______________________________________________________________________________
A visual impression of instantaneous particles configurations is provided in 
figure~\ref{fig:part_pos}, where slices of the domain in selected coordinate 
planes are displayed for three different Stokes numbers.
The typical particle distribution exhibits many voids, strongly correlated with 
high enstrophy, \cite{squeat,becetal}, intertwined with thin ``stretched'' 
regions where particles concentrate.
Clustering is specially manifest near $St_\eta=1$, see the mid panel of the 
figure in comparison with top and bottom ones which show more even spreading. 
The typical void dimension, as caught by the eye, is larger at our
largest Stokes number, $St_\eta=10$.  
The ballistic limit, where particles follow their trajectories with 
no significant influence from the fluid and the expected spatial distribution 
is homogeneous, is apparently still far away.
In the opposite extreme case, passive tracers are recovered for vanishing 
$St_\eta$, particles follow the fluid path, and, again, homogeneity is 
eventually restored.
In fact, clustering still takes place, though at smaller scales, for the 
smallest Stokes number we have considered, consistently with theoretical 
arguments~\cite{balfalfou,falfouste} and numerical 
simulations~\cite{becetal,reacol} for homogeneous and isotropic turbulence 
aimed at explaining droplets growth in clouds.

Despite of the mentioned similarities with isotropic flows, ours manifests 
specific features compelled by large scale anisotropy.
The shear-induced orientation is apparent from the bottom-left/top-right 
alignment of particles sheets in the shear plane $x-y$, see the right panels 
of figure~\ref{fig:part_pos}.
This behavior, clearly visible at $St_\eta=1.0$, is still discernible in 
the other two  cases.  
Since, apparently, the effect is strong and persistent, we are interested in 
putting forward suitable tools to evaluate the anisotropy of clustering. 
As we shall see, this is best done by extending in due form a line 
of analysis proved successful for the isotropic case.

The main statistical tool is the radial distribution function (RDF) of 
particle pairs $g(r)$ which is a function of radial distance $r$,
see e.g. \cite{suncol} where the RDF is dealt with for isotropic flows.
The RDF, sometimes called correlation function, is defined as
\begin{equation}
g(r)=\frac{1}{4\pi r^2} \frac{d N_r}{dr} \frac{1}{n_0} \, ,
\label{eq:g_r_sca}
\end{equation}
where $n_0 = 0.5 N_p(N_p-1)/V_0$ is the density of pairs in the whole volume 
$V_0$ and $N_r$  is the number of pairs in a ball ${\cal B}_r$ of radius 
$r$.

The concept is easily extended to anisotropic cases by considering the 
number of pairs $d \mu_r = \nu_r(r, {\bf \hat {r}}) d \Omega$ 
contained in a spherical cone of radius $r$, with axis along the direction 
${\bf \hat {r}}$ and solid angle $d \Omega$, see the sketch in the right panel 
of figure~\ref{fig:sketch}.
By this definition the number of pairs in the ball ${\cal B}_r$ is $N_r = \int_\Omega \nu_r d \Omega$, hence
$dN_r/dr = \int_\Omega d \nu_r/dr \, d \Omega$.
We define the Angular Distribution Function (ADF) as
\begin{equation}
	g(r,\hat{\bf r})=
        \frac{1}{r^2} \frac{d \nu_r}{dr} \frac{1}{n_0} \, ,
\label{eq:g_r_vec}
\end{equation}
which retains information on the angular dependence of the distribution. 
The RDF is the spherical average of the ADF
$g(r)=1/(4 \pi) \int_\Omega g(r,{\bf \hat {r}}) d\Omega$ and it is shown in 
figure~\ref{fig:radial_pdf} for a few particle populations.

The behavior of the RDF near the origin,
$g(r) \propto r^{-\alpha}$, is related to important geometrical features of
the spatial distribution. Specifically, 
${\cal D}_2  = 3 - \alpha$ is the so-called correlation dimension of the 
multi-fractal measure associated with the particle density, \cite{grassberger}.
A positive $\alpha$ indicates the occurrence of small
scale clustering. Its value is inferred from the  slope near the origin in the
log-log plots shown in figure~\ref{fig:radial_pdf}, see the scaling
behavior apparent in the range $r/\eta\in[.1:1]$.
From the figure, particles with $St_\eta \sim 1$ exhibit maximum accumulation,
i.e. the RDF diverges at a faster rate as $r$ is decreased,
see also \cite{shobal}.
The solid lines superimposed to the present data correspond to the scaling 
laws extracted at matching Stokes number by \cite{becetal} in isotropic 
conditions. The agreement between our data
and the isotropic ones is remarkable, showing that, even under strong shear, 
certain features of the clustering process may be universal.

Small scale clustering is not shared by heaviest particles. They show instead
the saturation of the correlation function to a constant value
$g(r) \simeq g_* > 1$ below a critical scale $\ell_c$,
see e.g. the open squares in figure~\ref{fig:radial_pdf} with 
$\alpha \simeq 0$.
This means that the number of pairs below $\ell_c$ is proportional to volume 
with an effective density $n_* = g_* n_0$ larger than its overall mean 
value $n_0$. 
By inspection of figure~\ref{fig:radial_pdf}, for the heaviest 
particles (squares), the saturation occurs at $\ell_c \simeq 10 \eta$.
The interpretation is that eddies with a time scale sufficiently smaller then 
particle Stokes time do not influence the clustering process. 
The saturation scale should then correspond to the size $\ell_{min}$ of the 
smallest eddies able to aggregate particles.
As an order of magnitude estimate, our data are consistent with the results  
given by \cite{yoshi_goto} for the inertial range of isotropic turbulence, where
the authors find $\ell_{min}/\eta = \left(St_\eta/\beta_{min}\right)^{3/2}$ 
with $\beta_{min}\simeq2$. Our data match this estimate also for the run at 
$St_\eta=5$, where the saturation occurs close to the Kolmogorov scale.
For lighter particles, clustering keeps on going below the Kolmogorov 
length and its lower limit cannot be interpreted by arguments which, tuned
by experiment, are nevertheless taken from inertial range theory.

Concerning the large scale behavior, heavier particles apparently begin to 
show accumulation at larger scales.
Each population is uniform with  $g(r) \simeq 1$ at very large $r$.
This trend is followed by the lightest particles  down to
$5\,\eta$ below which they begin to follow the power-law.
The RDF starts deviating from the uniform distribution--$g(r) \simeq 1$-- 
at a scale $\ell_{max}$ which increases monotonically with the Stokes number, 
see. e.g. the range $r/\eta\in[20:80]$. 
This is consistent with intuition, since the particle Stokes time 
progressively matches the eddy-turnover time of larger turbulent eddies.
The data again agree reasonably well with the results of 
\cite{yoshi_goto}, which identify the range of time scales relevant to 
clustering as $\beta_{max} < \tau_p/\tau(\ell) < \beta_{min} $ where 
$\tau(\ell)\propto \epsilon^{-1/3} \ell^{2/3}$ is the eddy turn over time at
scale $\ell$. 
In terms of lengths, clustering starts to occur at
$\min{\left( \ell_0,\ell_{max}\right) }$ where $\ell_0$ is the integral scale 
and $\ell_{max}/\eta = \left(St_\eta/\beta_{max}\right)^{3/2}$ 
($\beta_{max}\simeq0.1$). 
Using this estimate, our heaviest particles are expected to begin accumulating
at the integral scale $\ell_0/\eta\simeq 135$ while, e.g., the lightest 
ones at $\ell_{max}/\eta \simeq 1$, which is not too far from the value 
$5\,\eta$ we infer from the plot.

\begin{figure}
\begin{center}
\includegraphics[width=.45\textwidth]{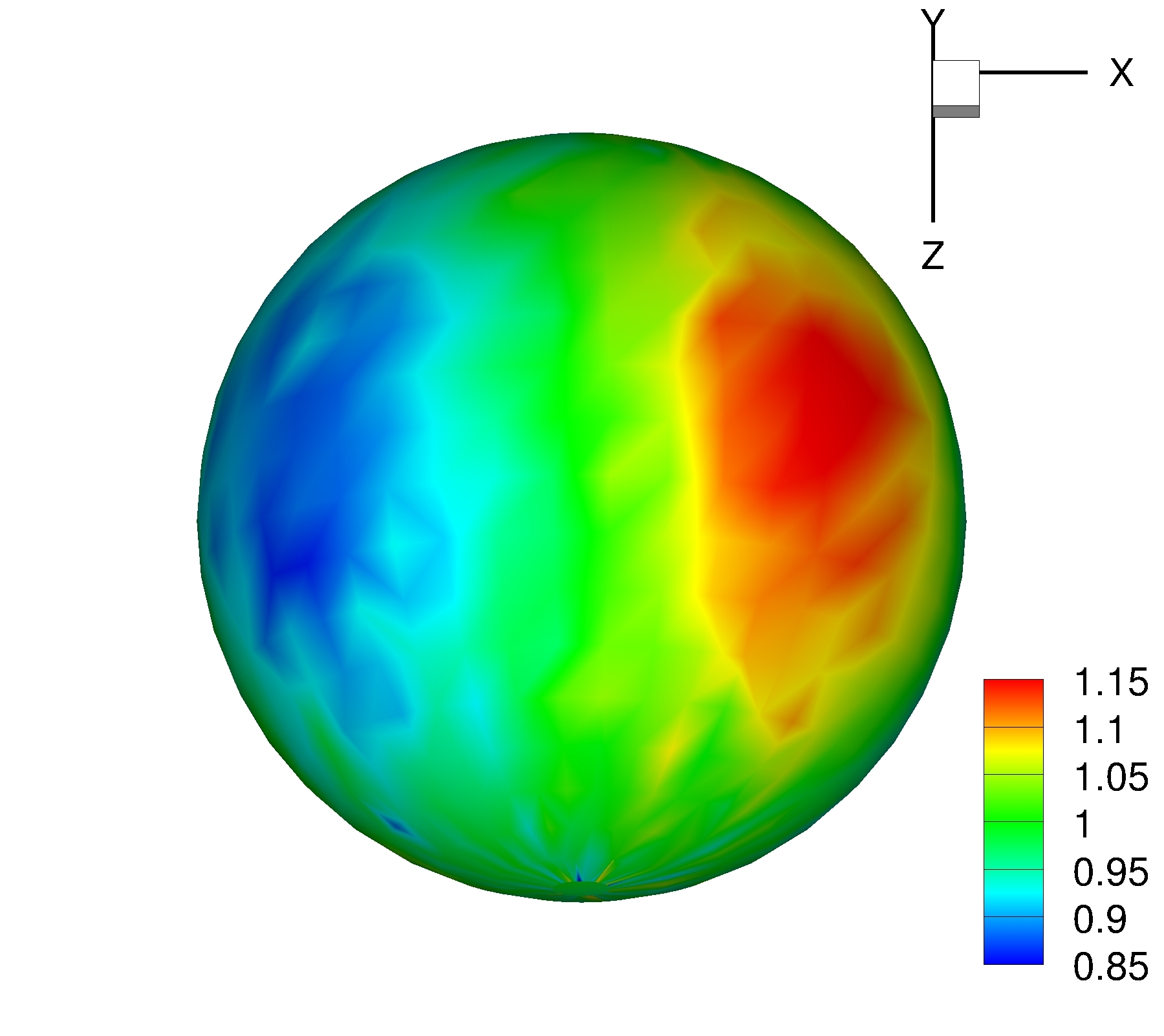}
\includegraphics[width=.45\textwidth]{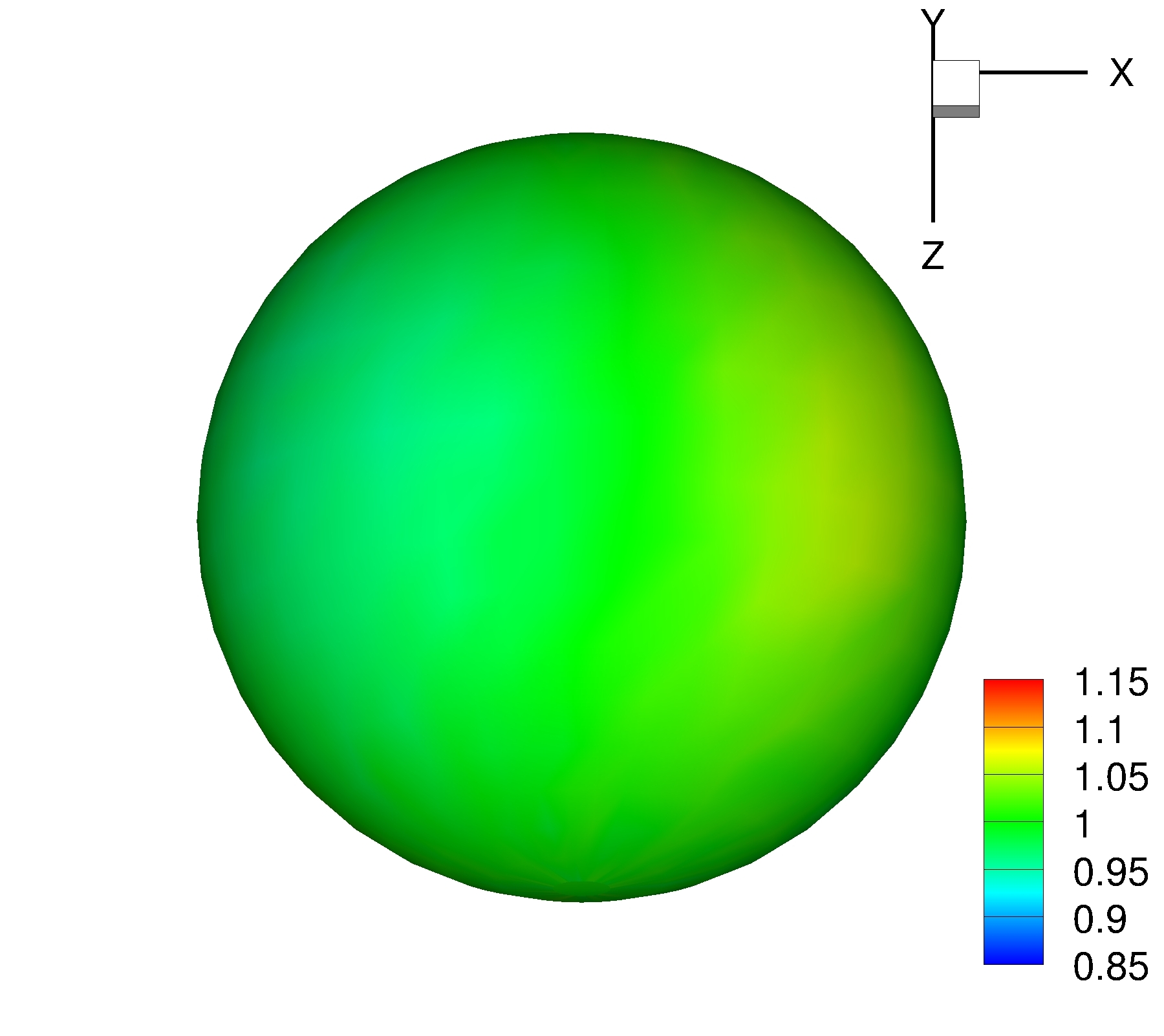} \\
\includegraphics[width=.45\textwidth]{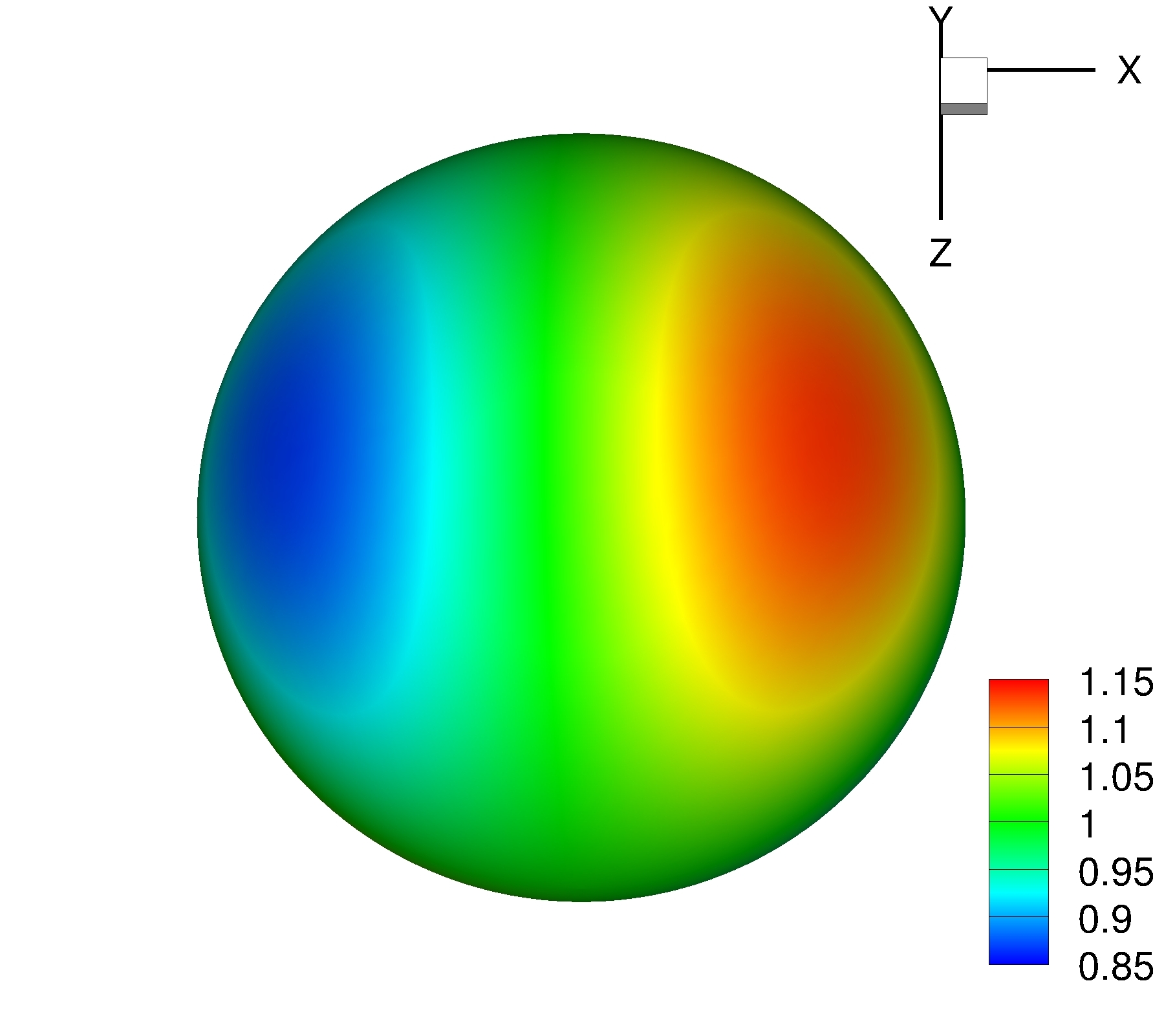}
\includegraphics[width=.45\textwidth]{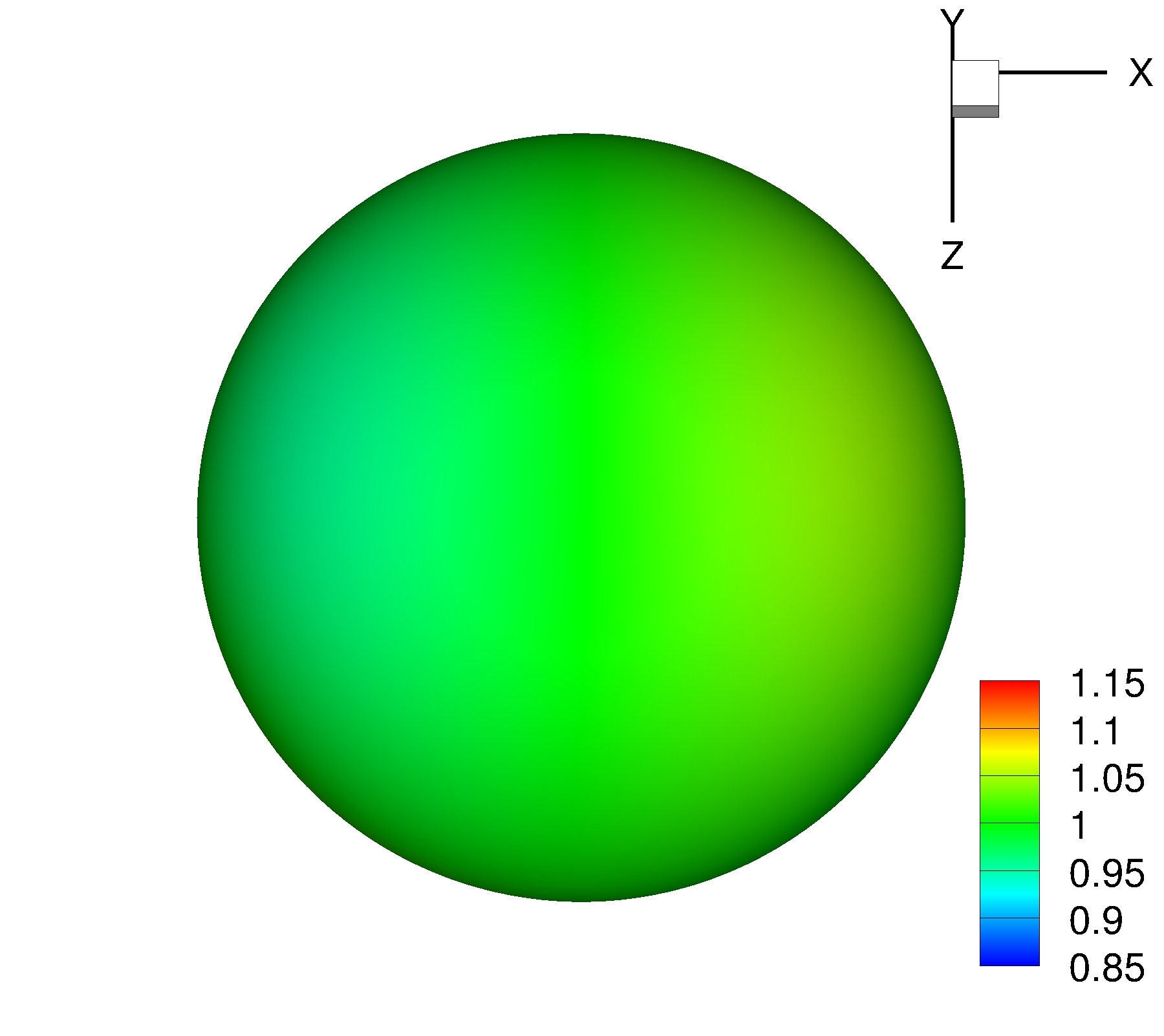}
\end{center}
\caption{Top panels: Angular Distribution Function (ADF) giving the probability per unit 
solid angle to find a couple of particle at fixed distance $|\bf r|$. Left panel
ADF computed at separation $4\eta$; right panel $35\eta$. Data for $St_\eta=1$.
Bottom panels: Estimate of ADF by using only the isotropic sector and the
$j=2$ sector.
         \label{fig:ball_vera} }
\end{figure}

The RDF quantitatively confirms the overall impression gained from the 
visualizations of figure~\ref{fig:part_pos}. 
The strong anisotropy apparent in those
plots, however, needs a description in terms of the more complete ADF
shown in figure \ref{fig:ball_vera} as a contour plot on the unit 
sphere for two separations $r$ and particles with $St_\eta=1$. 
The ADF has been normalized with its average on the unit sphere, $g(r)$, 
in order to compare the relative anisotropy content of different scales.
From figure \ref{fig:ball_vera}, as separation decreases 
(i.e. moving  from the right, $r = 35 \eta$, to the left panel, $r = 4 \eta$), 
the normalized ADF exhibits preferential clustering in directions consistent 
with those observed in the visualizations of figure~\ref{fig:part_pos}. 

The ADF allows for a systematic evaluation of anisotropy in particle clustering.
For given separation $r$, its angular dependence can be resolved in terms
of spherical harmonics,
%_______________________________________________________________________________
%_______________________________________________________________________________
\begin{equation}
\label{so3_dec}
g(r,{\bf \hat{r}})= \sum_{j=0}^\infty \sum_{m=-j}^j \, g_{jm}(r) \,
Y_{jm}(\hat{\bf r}) \ .
\end{equation}
%_______________________________________________________________________________
%_______________________________________________________________________________
In this notation, the classical RDF $g(r)$ is
the projection of the ADF on the isotropic sector $j=0$, namely
%_______________________________________________________________________________
%_______________________________________________________________________________
\begin{equation}
\label{g_00}
g_{00}(r)\equiv g(r) =\int_\Omega g(r, {\bf \hat{r}}) \, 
Y_{00}(\hat{\bf r}) \, d\Omega.
\end{equation}
%_______________________________________________________________________________
This decomposition amounts in projecting our function on 
orthogonal sub-spaces invariant under rotations. Each sub-space is labeled by 
the index $j$ and it is spanned by $2j+1$ base elements $Y_{jm}$.
Growing levels of anisotropy are checked by increasing $j$, 
consistently with its geometrical meaning as number of zero crossings
of $Y_{jm}$.

\begin{figure}
\begin{center}
\includegraphics[width=.9\textwidth]{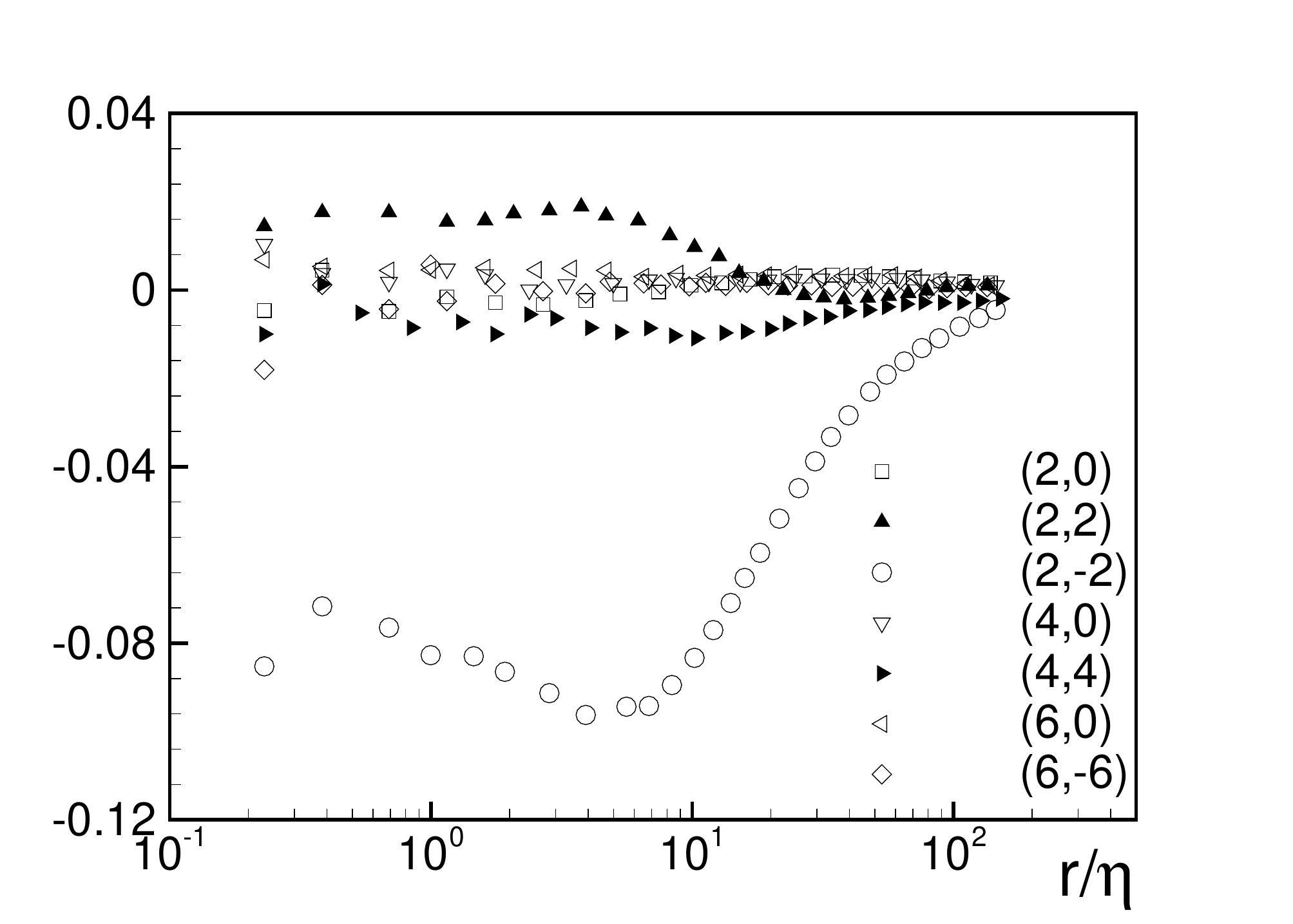}
\end{center}
\caption{Projection of ADF on the different anisotropic
sectors of spherical harmonics 
normalized by the projection of the isotropic sector (RDF) as a function
of separation. Data for $St_\eta=1$.
         \label{fig:y_ij} }
\end{figure}

Figure \ref{fig:y_ij} shows the normalized projections on sectors $j=2,4,6$,
i.e. the normalized amplitudes $g_{jm}/g_{00}$, for the particle population
with $St_\eta = 1$.
For given $j$ most modes are negligible and the figure reports 
only those with significant signal level. 
The $(2,-2)$ mode provides the most significant contribution to the anisotropic
component of $g(r,{\bf \hat{r}})$.
The corresponding spherical harmonics, $Y_{2-2}$, roughly selects the 
intensity of the signal along the principal direction of
the mean deformation tensor which corresponds to maximum straining 
and is inclined $45^\circ$ in the mean flow plane. $Y_{2-2}$ is negative in 
the first and third quadrant and positive in the others, thus explaining the 
negative sign of $g_{2-2}$ in figure~\ref{fig:y_ij}.
This description well captures the alignment of most thin particle clusters 
observed in the mid panel of figure \ref{fig:part_pos}.

The signal content rapidly decreases with the order of the sector $j$
and the ADF is satisfactorily reconstructed--see 
decomposition~(\ref{so3_dec})--using only the first few sectors with 
$j \le 2$ as shown in bottom panels of figure \ref{fig:ball_vera}.
We conclude that the probability of finding two particles at
separation $r=4 \eta$ along the main straining direction is $30\%$ higher than 
in the perpendicular direction in the mean flow plane.

As discussed, the ADF provides a quantitative account of the anisotropy 
induced by the fluid velocity field on the disperse phase. As shown
below, it can be effectively used to parameterize the level of anisotropy 
through the scales in terms of the Stokes number.

The plot of figure~\ref{fig:y_2_2} gives the normalized amplitude of the most 
energetic anisotropic mode in absolute value--$|g_{2-2}|/g_{00}$--for our 
set of Stokes numbers $St_\eta=10,5,1,0.5,0.1$ ranging from heavy to light 
particles.  
Focusing on the heaviest particles, $St_\eta=10,5$, 
the relative amplitude of the strongest anisotropic mode first increases
towards the small scales to reach a maximum at $r\sim \ell_c$. Below this scale
the anisotropy level decreases, until the very small scales become 
essentially isotropic. Connecting this result with the previous discussion
concerning the saturation of the RDF $g_{00}$,
we conclude that the heaviest particles show a regular concentration at
scales smaller than $\ell_c$ where the distribution recovers isotropy.

Particles with smaller Stokes numbers behave in an entirely different way.
The anisotropy, as measured by the ratio $g_{2-2}/g_{00}$, substantially 
increases to saturate at small scales close to Kolmogorov length. 
It keeps an almost constant value below 
$\eta$. In other words, the clustering process maintains its anisotropic 
features even below the dissipative scale for sufficiently small Stokes number 
particles. Remind that the overall clustering process described by $g_{00}$ is 
here characterized by a singular exponent $\alpha$.  The saturation observed 
on the ratio $g_{2-2}/g_{00}$ implies that the dominating anisotropic
contribution inherits the same behavior, $g_{2-2} \propto r^{-\alpha}$.

\section{Final comments} \label{comments}

We have provided evidence that large scale shear induces preferential 
orientation on the patterns a disperse phase of small inertial particles
forms in turbulence. The effect is indirect: The shear imprints anisotropy 
on velocity fluctuations which, in turn, arrange particle configurations in 
directionally biased clusters.

Recently it became increasingly clear that the multi-scale nature of the 
velocity field is crucial in explaining most features of particles 
distributions.
Typically, a range of eddies exists able to break spatial homogeneity
of particles configurations.
The particles segregate and their pattern shows local concentrations
and voids on length scales correlated with those of the inducing eddies.
The range of scales of the inducing eddies is determined by the Stokes time of
the particles, and moves from integral down to Kolmogorov length with 
reducing the relaxation time.
For light particles, clustering reaches down Kolmogorov scale, leading 
to a singularity in the radial distribution function.
The exponent of the singularity is a measure of its intensity.
All these features, originally found in experiments and numerical simulations 
of isotropic turbulence, are also present in our anisotropic fields. 
They are thus generic aspects of the clustering process which seem 
independent of the specific geometry of the forcing.

At a qualitative level, the specific characteristics of clustering under 
anisotropic advection consists of preferential orientation of the particle 
patterns.
In order to quantify the new scale-dependent features added by the non-trivial 
geometry of the forcing, we have introduced the concept of angular 
distribution function.
It can be understood as a generalization of the previous radial distribution
function, to which it reduces by averaging on the unit sphere, i.e. by 
performing the isotropic projection.
This quantitative tool has led to our most unexpected finding: 
The advecting field anisotropy, known to be confined to the large scales,
affects the singular, small scale, clustering process.
In fact, anisotropy results into a strong directionality of the probability to 
find a couple of particles at viscous scale separation, with $30 \%$ variations 
on the solid angle easily observed.

Technically, for Stokes number order unity, the anisotropic component of the 
angular distribution function diverges at small scales with singularity 
exponent comparable to that found in the isotropic projection (RDF).
For very small Stokes number, we cannot even exclude that the singularity 
exponent of the strongest anisotropic sector may even exceed
that of the radial distribution function. 
Conversely, heavy particles appear to preferentially concentrate on finite 
sized patches endowed with a range of multi-scale and shear oriented features, 
with finest scales more or less evenly and isotropically distributed.

The geometrical properties  of patterns of inertial particles differ 
considerably from that one could naively guess from velocity fluctuations.
Recent findings, extending somehow a number of previous results on shear
induced anisotropy
\cite{Corrsin,uberoi,george,antonia,warshe}, show that velocity 
fluctuations manifest two neatly distinct ranges, one dominated by the 
production of turbulent kinetic energy above the shear scale $L_s$, the other,
corresponding to the classical inertial range of Kolmogorov theory below.
In the two ranges, velocity fluctuations display different isotropy recovery
rates, a smaller one in the production range, a larger one in the inertial 
transfer range between $L_s$ and the viscous scale $\eta$, 
\cite{casguajacpiv,jaccastalalf}. 
Actually, concerning the velocity field, isotropy is matter of fact always 
recovered at 
dissipative scales, provided the scale separation $L_s/\eta$ is large enough, 
i.e. the local Reynolds number (Taylor-Reynolds number, or, equivalently 
for wall bounded flows the wall normal distance in inner units, 
$y^+ = y \sqrt{\tau_w/\rho}/\nu$, with $\tau_w$ the wall shear stress) is 
sufficiently large.  This condition is violated in an essential way 
only very close to solid walls in wall-bounded turbulence (small $y^+$).

The anisotropy in the particle configurations depends strongly on the
properties of the advecting velocity field. 
However, despite of the isotropy recovery of the velocity field, isotropy may 
never be recovered in the small scales of the clusters, as it happens at small 
Stokes number.  Actually, in the range from $L_s$ to $\eta$ clustering of 
light inertial particles shows a substantial increase of directionality.

\begin{figure}
\begin{center}
\includegraphics[width=.9\textwidth]{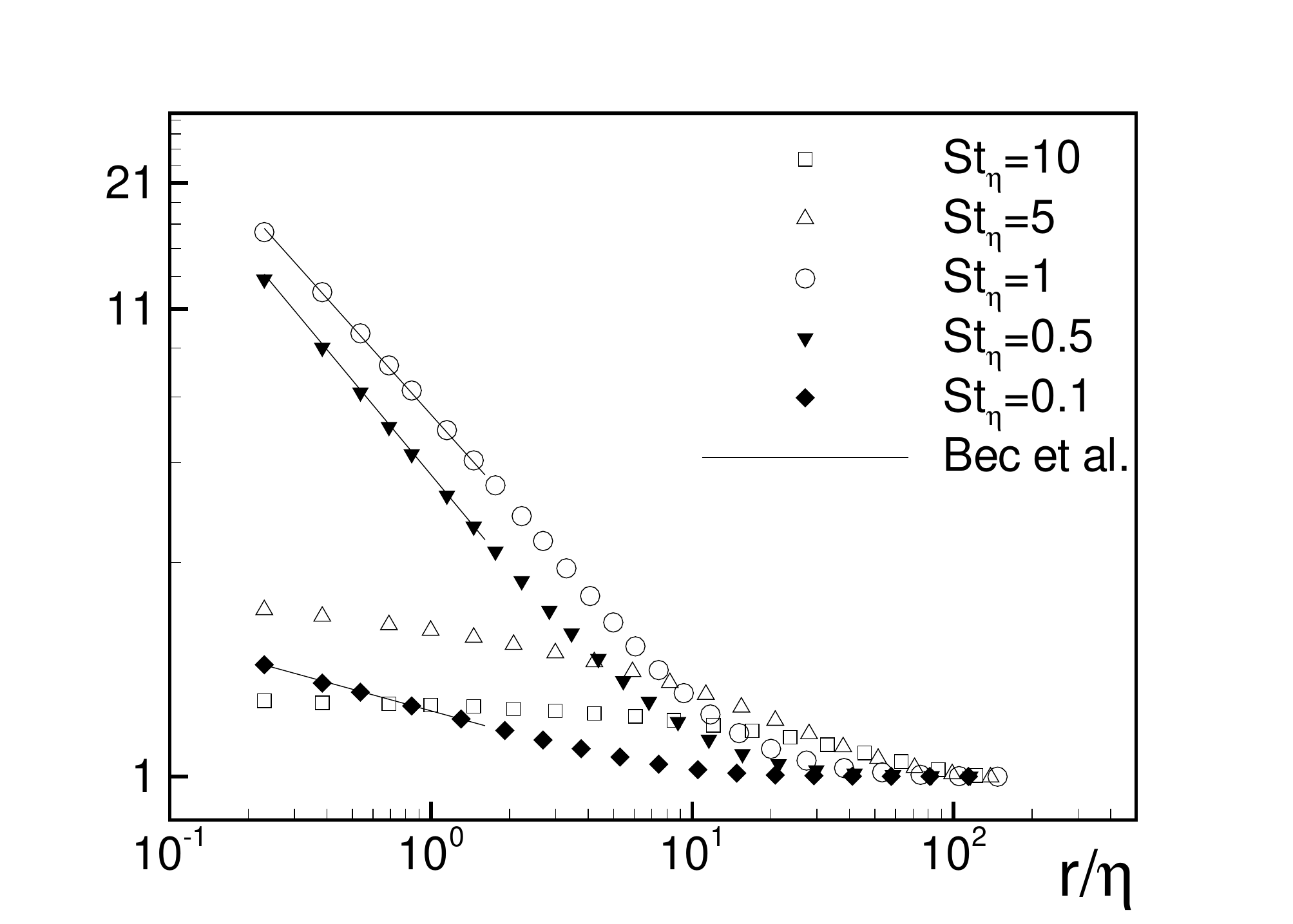}
\end{center}
\caption{Radial distribution function $vs$ separation,
for different Stokes number.
         \label{fig:radial_pdf} }
\end{figure}

\begin{figure}
\begin{center}
\includegraphics[width=.9\textwidth]{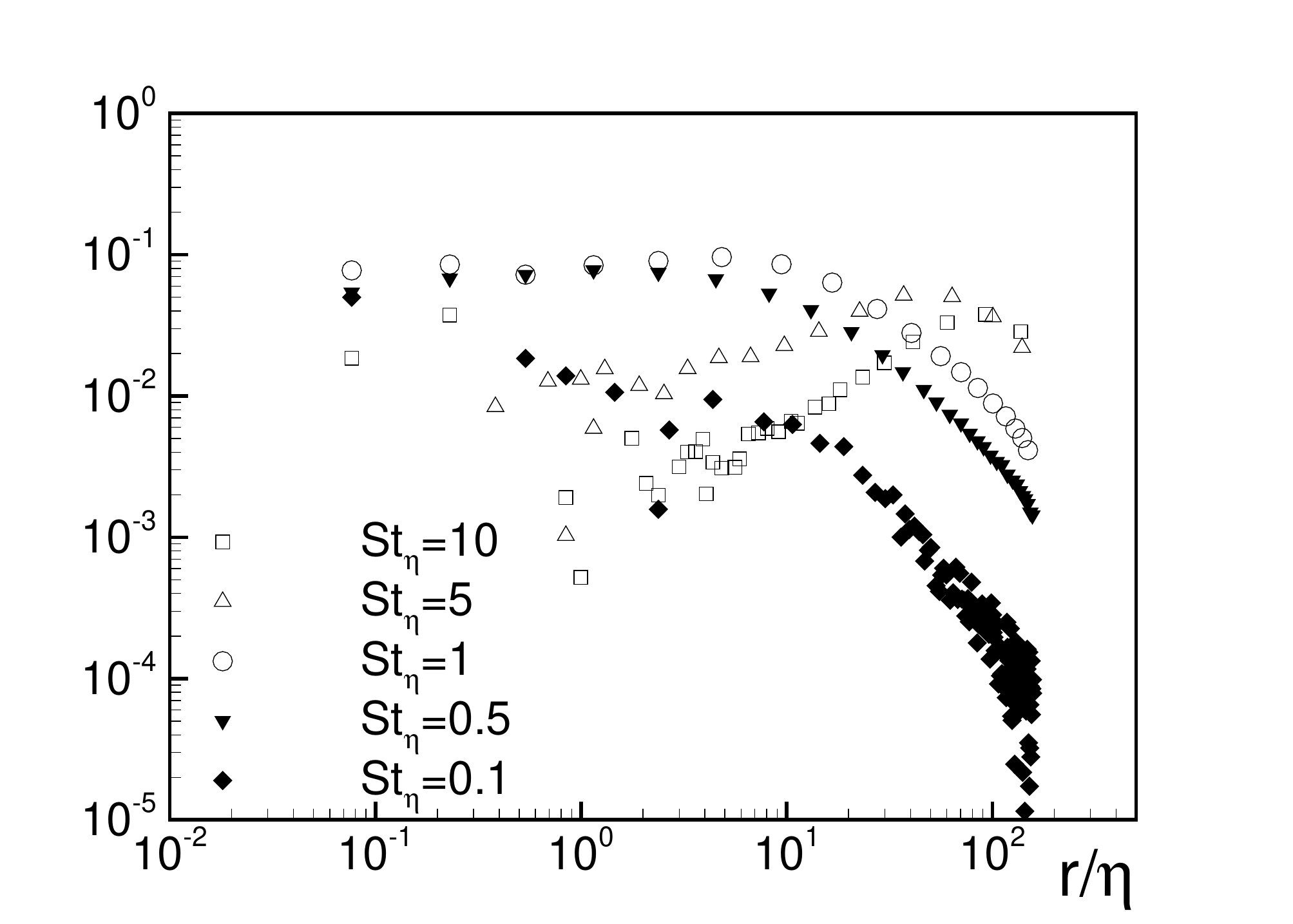}
\end{center}
\caption{Ratio between the most energetic anisotropic sector $(2,-2)$
normalized by isotropic sector as a function
of separation for different Stokes number.
         \label{fig:y_2_2} }
\end{figure}

From our results at moderate Reynolds number one can conjecture the behavior 
of the clusters at high Reynolds numbers.
In principle the intermittency of the turbulent field may induce a dependence 
on Reynolds number.
In fact, as shown in \cite{becetal} for isotropic flows, the fractal properties
of particle distributions depend at most weakly on Reynolds number and 
strongly on the Stokes number. 
For given geometry of the external forcing, i.e. fixing the integral and the 
shear scale, and at given particle Stokes time, the increase of the Reynolds 
number corresponds to increasing the Stokes number based on Kolmogorov time, 
$St_\eta$. Given the weak dependence on Reynolds number, 
this is somehow equivalent to reading figs.~\ref{fig:radial_pdf} 
and \ref{fig:y_2_2} by successively moving from light to heavy particles. 
Along the process, we infer the saturation at small scales of the radial 
distribution function ($St_\eta \gg 1$, fig.~\ref{fig:radial_pdf}) and the
small scale isotropy of the clusters (fig.~\ref{fig:y_2_2}). 
Clustering is confined to the intermediate scales where it shows high levels 
of anisotropy.
This means that, in the limit of large Reynolds number, any particles 
population should be organized in anisotropic finite-sized patches which are 
eventually uniform and isotropic in their finest scales.
We stress once more  that an intermediate range of scales always exists, 
however, where a multi-scale aggregation process takes place with a 
substantial directionality of the clusters.

On the other hand, at finite Reynolds number, sufficiently small 
particles--small relaxation time--will always show small scale clustering, in 
the sense of a singularity in the radial distribution function. 
In this case anisotropy may persist below Kolmogorov scale, as described by 
the angular distribution function at small separations for the lightest 
populations in our simulations.
This finite Reynolds number effect becomes extremely important in the near 
wall region of turbulent wall bounded flows, where the local Reynolds number 
constructed with the distance from the wall is not huge and the velocity field 
is meanwhile strongly anisotropic. 

The issuing anisotropy of the fine scales of particles clusters will then have 
a significant impact on phenomena of collision, aggregation of dusts into 
larger particles, evaporation/condensation rate of droplets in pipe lines and 
a number of other physically and technologically significant contexts.

A final comment concerns the extension of the present results to the near wall 
region of wall bounded flows. 
As shown by recent results on the scale-by-scale statistics of the velocity 
field, e.g. energy transfer across scales, spectral distribution of turbulent 
kinetic energy production, intermittency and anisotropy, the homogeneous shear 
flow reproduces the essential features observed in the wall region, despite of 
significant differences in the large scale geometry of the two systems.
However, particles in the wall region are strongly affected by turbophoresis
which is a predominant effect associated with inhomogeneity. In wall flows 
particle segregation is controlled by the two concurrent processes of small 
scale anisotropic clustering and of accumulation at the wall. Clearly, the 
focus of the present paper is on the former one, leaving the combined analysis 
of the two effects for future investigations.
%_______________________________________________________________________________

%___________________________________________________________________________
\clearpage
\newpage
%____________________________________________________________________________________________
%%
%% Figure 1: nomenclature
%%
%____________________________________________________________________________________________
%_______________________________________________________________________________
\clearpage
\newpage
%____________________________________________________________________________________________
%%
%% Figure 1-bis: interpolation schemes
%%
%____________________________________________________________________________________________
%_______________________________________________________________________________
\clearpage
\newpage
%_______________________________________________________________________________
%%
%% Figure 2: particle position
%%
%_______________________________________________________________________________
\clearpage
\newpage
%_______________________________________________________________________________
%%
%% Figure 3: radial pdf
%%
%_______________________________________________________________________________
\clearpage
\newpage
%_______________________________________________________________________________
%%
%% Figure 4: ball_vera e ball_ricostruita
%%
%_______________________________________________________________________________
\clearpage
\newpage
%_______________________________________________________________________________
%%
%% Figure 6: Y_mj/Y_00 St=1
%%
%_______________________________________________________________________________
\clearpage
\newpage
%_______________________________________________________________________________
%%
%% Figure 8: Y_2_2/Y_00 different St
%%
%_______________________________________________________________________________
%___________________________________________________________________________
\end{document}